\documentclass[fleqn,10pt]{SelfArx} 
\usepackage[numbers,sort&compress]{natbib}
\usepackage[font={small,singlespacing}]{caption}
\usepackage{setspace}
\usepackage{hyperref}
\usepackage{lettrine}

\definecolor{color1}{RGB}{39,45,102} 
\definecolor{color2}{RGB}{240,240,240} 

\newlength{\tocsep} 
\usepackage{times}

\JournalInfo{April 2013} 
\Archive{} 

\PaperTitle{Limited communication capacity unveils strategies for human interaction} 

\Authors{Giovanna Miritello\textsuperscript{1,2}, Rub\'en Lara\textsuperscript{2}, Manuel Cebri\'an\textsuperscript{3,4}, Esteban Moro\textsuperscript{1,5,}*} 
\affiliation{\textsuperscript{1}\textit{Departamento de Matem\'aticas \& GISC, Universidad Carlos III de Madrid, 28911 Legan\'es, Spain}} 
\affiliation{\textsuperscript{2}\textit{Telef\'onica Research, 28050 Madrid, Spain}} 
\affiliation{\textsuperscript{3}\textit{National Information and Communications Technology Australia, Melbourne, Victoria 3010, Australia}}
\affiliation{\textsuperscript{4}\textit{Department of Computer Science and Engineering, University of California at San Diego, La Jolla, CA 92093, USA}}
\affiliation{\textsuperscript{5}\textit{Instituto de Ingenier\'{\i}a del Conocimiento, Universidad Aut\'onoma de Madrid, 28049 Madrid, Spain}}
\affiliation{*\textbf{Corresponding author}: emoro@math.uc3m.es} 

\Keywords{\footnotesize Social networks, communication capacity, social strategy, information diffusion} 
\newcommand{\keywordname}{Keywords} 

\Abstract{ Social connectivity is the key process that characterizes the structural properties of social networks and in turn processes such as navigation, influence or information diffusion. Since time, attention and cognition are inelastic resources, humans should have a predefined strategy to manage their social interactions over time. However, the limited observational length of existing human interaction datasets, together with the bursty nature of dyadic communications have hampered the observation of tie dynamics in social networks. Here we develop a method for the detection of tie activation/deactivation, and apply it to a large longitudinal, cross-sectional communication dataset ($\approx$ 19 months, $\approx$ 20 million people). Contrary to the perception of ever-growing connectivity, we observe that individuals exhibit a finite communication capacity, which limits the number of ties they can maintain active. In particular we find that men have an overall higher communication capacity than women and that this capacity decreases gradually for both sexes over the lifespan of individuals (16-70 years). We are then able to separate communication capacity from communication activity, revealing a diverse range of tie activation patterns, from stable to exploratory. We find that, in simulation, individuals exhibiting exploratory strategies display longer time to receive information spreading in the network those individuals with stable strategies. Our principled method to determine the communication capacity of an individual allows us to quantify how strategies for human interaction shape the dynamical evolution of social networks.
}

\begin{document}

\flushbottom 

\maketitle 


\thispagestyle{empty} 

\setstretch{0.92}

\lettrine[nindent=0pt, lines=2]{M}{any} different forces govern the evolution of social relationships making them far from random. In recent years, the understanding of what mechanisms control the dynamics of activating or deactivating social ties have uncovered forces ranging from geography to structural positions  in the social network ({\it e.g.} preferential attachment, triadic closure),  to homophily~\cite{uzzi}. These finding are pervasive in empirical analyses across cultures, communication technologies and interaction environments~\cite{burt,martin,hidalgo,raeder,kossinets2,crandall,romero,wang2012cooperation,apicella2012social,rand2011dynamic}.

However, the incorrect assumption that time, attention and cognition are elastic resources has blurred the study of how individuals manage their social interactions over time \cite{wu2007novelty,hodas2012visibility,backstrom2011center}. Understanding such social strategies is not only of paramount importance to make progress in the characterization of human behavior, but also to improve our current description of social networks as evolutionary objects against the (aggregated) ever-growing or static pictures of the social structure. 

Several reasons have hampered the observation of tie activation/deactivation dynamics in social networks at large scale: on the one hand, studies of diffusion based on datasets from pre-electronic eras have safely assumed that tie activation/deactivation is a much slower process than interactions within a tie, and thus their dynamics might be safely neglected \cite{hagerstrand1968innovation,rogers1995diffusion,christakis2007spread}. However, the current ability to communicate faster and further than ever accelerates tie dynamics in an unprecedented manner to the point that tie activation/deactivation may rival in time with processes like information spreading. On the other hand, available data about how ties form or decay were restricted to egocentric, small social networks and/or short periods of time which made it difficult to assess the universality of the results obtained and their extension to other situations \cite{raeder}. Finally, although in some online social networks there are explicit rules for the establishment of social ties, in most cases activity is the only way to assess the existence or not of the tie \cite{huberman,przemek}. Online social networks are plagued with this problem due to the cheap cost of maintaining ``friends'' which are in fact already deactivated relationships \cite{sibona2011unfriending}. However, using activity as proxy for tie presence is a problem in most communication channels like mobile phone calls, emails, electronic social networks etc., since tie activity is very bursty \cite{barabasiburst} and so far there is no clear method to discriminate those social ties that are already inactive from large-inter even times within active relationships \cite{song2012}.

\begin{figure*}[t]
\centerline{\includegraphics[width=.7\textwidth]{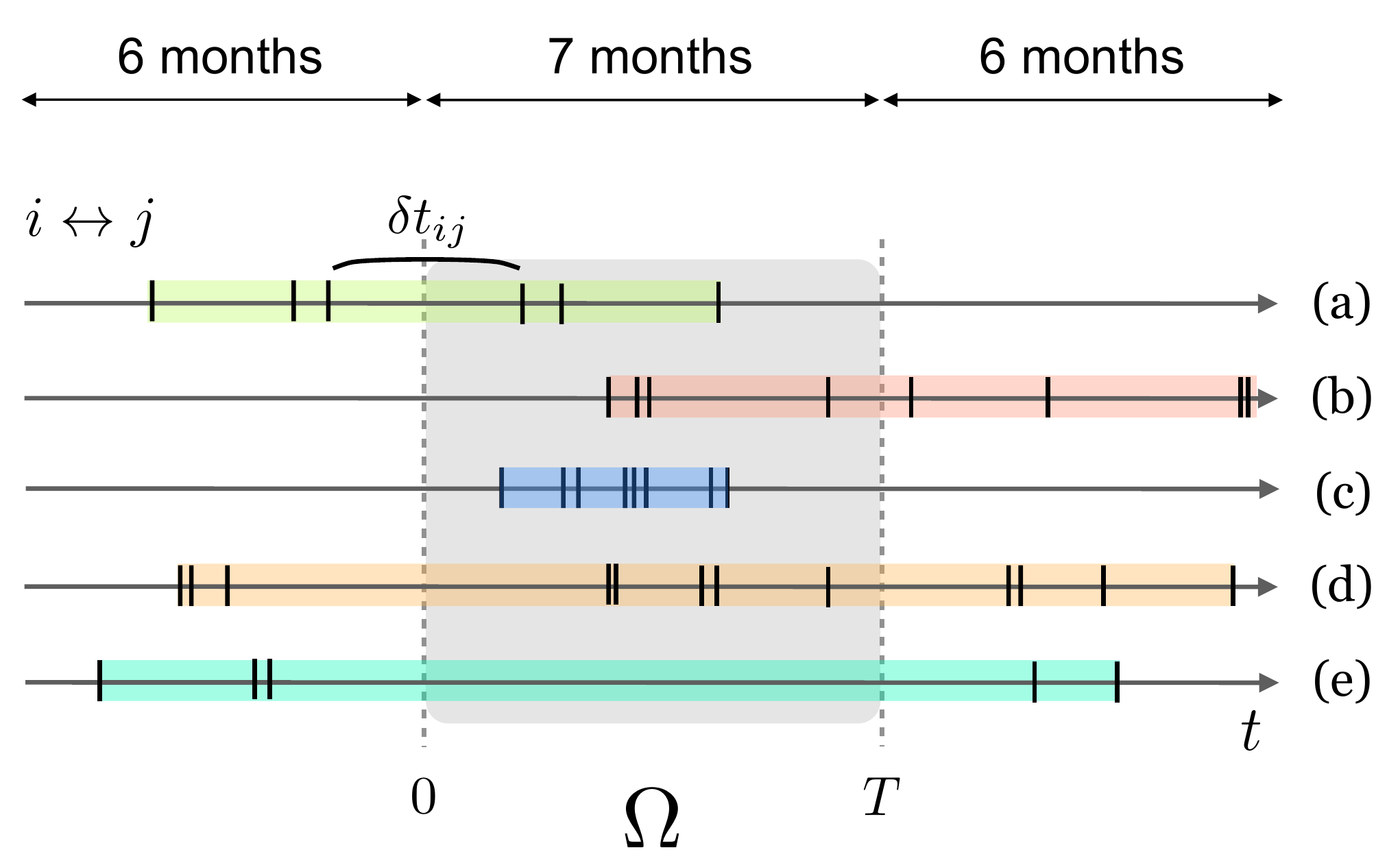}}
\caption{{\bf Detection of tie activation/deactivation:} Schematic view of the time intervals considered in our database and the different situations of tie activation/deactivation and the interplay between the tie communication patterns and tie activation/deactivation for a given observation time window $\Omega$ of length $T = 7$ months  (shadowed area). Each line refers to a different tie while each vertical segment indicates a communication event between $i\leftrightarrow j$ and $\delta t_{ij}$ is the inter-event time in the $i\leftrightarrow j$ time series. \label{fig1}}
\end{figure*}

\section{Detection of tie activation/deactivation}
To study the formation and decay of communication ties, we study the Call Detail Records (CDRs) from a single mobile phone operator over a period of $19$ months. The data consists of the anonymized voice calls of about $20$ million users that form $700$ million communication ties. After filtering out all the incoming or outgoing calls that involve other operators, we only consider users that are active across the whole time period and retain only ties which are reciprocated. We refer to {\em SI Sections A \& H}  for further details about the processing and the sampling of the datasets and for the comparison with another (smaller) database of Facebook communication through wall posts.

In most studies of communication networks a tie is assumed to be present if it shows any activity in the observation window \cite{onnela}. However, since communication is bursty \cite{barabasiburst}, large inter-event times between interactions are likely and thus they might be unobserved or mistaken as tie decay or formation, specially if the observation window is short (see Fig.\,\ref{fig1} and {\em SI Section A}). For example, in our call database we find that the average time between tie communication events is $\langle \delta t_{ij}\rangle = 14$ days (with $\sigma = 18$ days) and thus we might get spurious effects if the observation window is of the order of months, as repeated interactions may fall outside the observation window \cite{blondel}.

To overcome this we propose a different method to asses whether a tie has been activated/deactivated in the observation window $\Omega$. The method is based on the {\it observation} of tie activity in a time window before/after $\Omega$: if tie activity is observed in the 6 months before $\Omega$ then it is considered an old tie [cases (a) and (d) in Fig.~\ref{fig1}]; on the other hand, if activity is observed in the 6 months after $\Omega$ we will assume that the tie persists [cases (b) and (d) in Fig.~\ref{fig1}]. In any other case, we will consider that the tie is activated and/or deactivated in $\Omega$ [cases (a), (b) and (c) in Fig.~\ref{fig1}]. Of course, it is possible that even if there is no communication before/after the observation window, the tie is still active after/before our database. This would require that the tie has an inter-event time $\delta t_{ij}$ bigger than 7 months, i.e. case (e) in Fig.~\ref{fig1}. However, in our database, only 3.5\% of the links have such a long inter-event time which validates the accuracy of our definition of tie activation/deactivation. See {\em SI Section B} for details on our discrimination method.

\begin{figure*}[t]
\centerline{\includegraphics[width=0.6\textwidth]{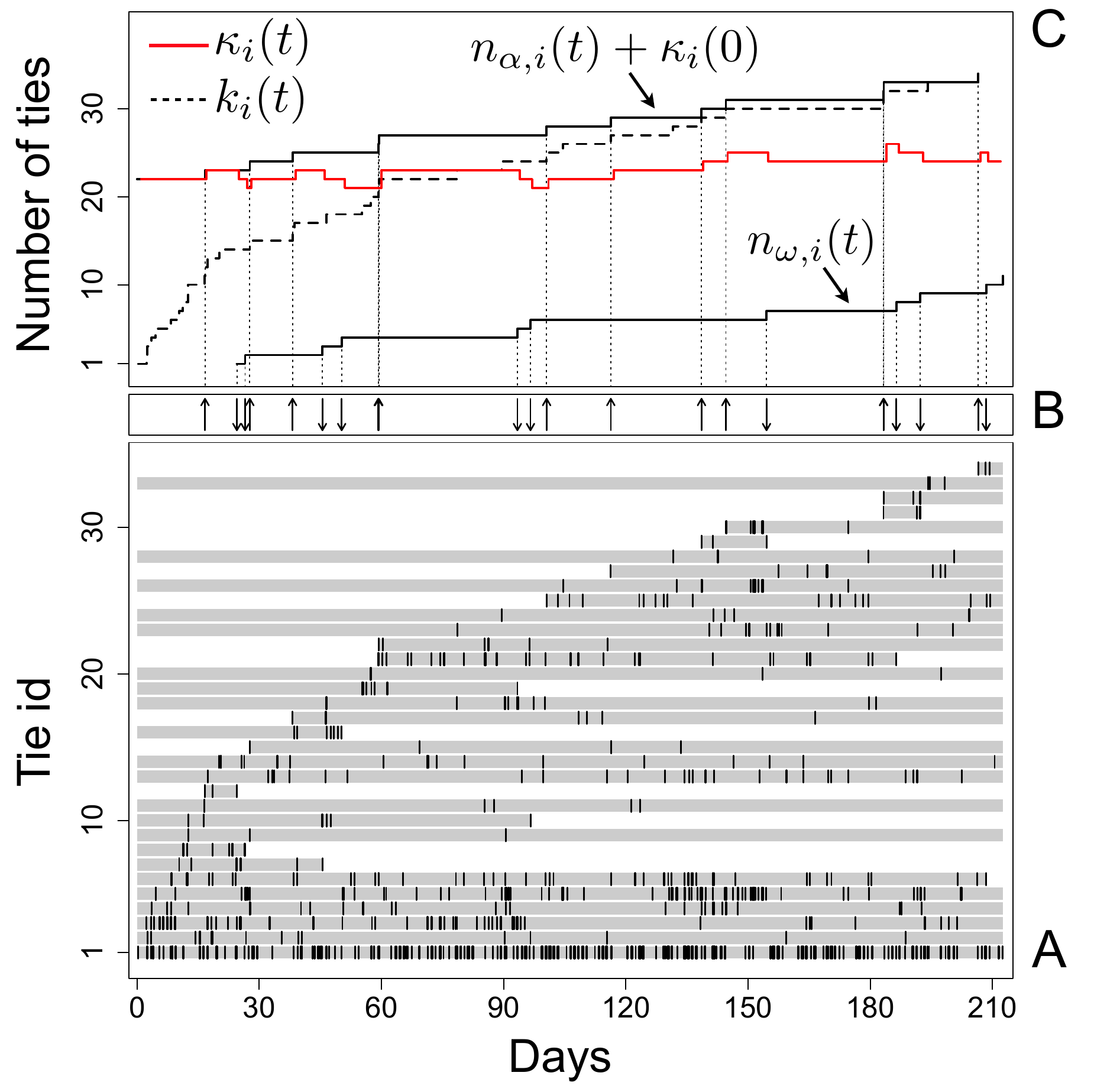}}
\caption{{\bf Communication capacity and evolution of activity:} Panel (A) shows the communication events of a given individual in our database with all her neighbors in the observation window $\Omega$. For each tie id, a vertical line represents a call with the corresponding neighbor. Grey horizontal rectangles are drawn from the first to the last observed communication event in each tie, considering also events before and after $\Omega$. Panel (B) shows vertical up/down arrows for each tie activation/deactivation events detected within $\Omega$. Using those events, panel (C) shows the aggregated number of active ties as a function of time $\kappa_i(0)+n_{\alpha,i}(t)$ and the aggregated number of deactivated ties $n_{\omega,i}$. Dashed line is the apparent growth in the social connectivity $k_i(t)$ obtained by the cumulative number of observed activity in ties up to some time, while red line is the number of active connections at a given instant $\kappa_i(t)$. \label{fig2}}
\end{figure*}

\section{Communication capacity and activity}
\begin{figure*}
\centerline{\includegraphics[width=\textwidth]{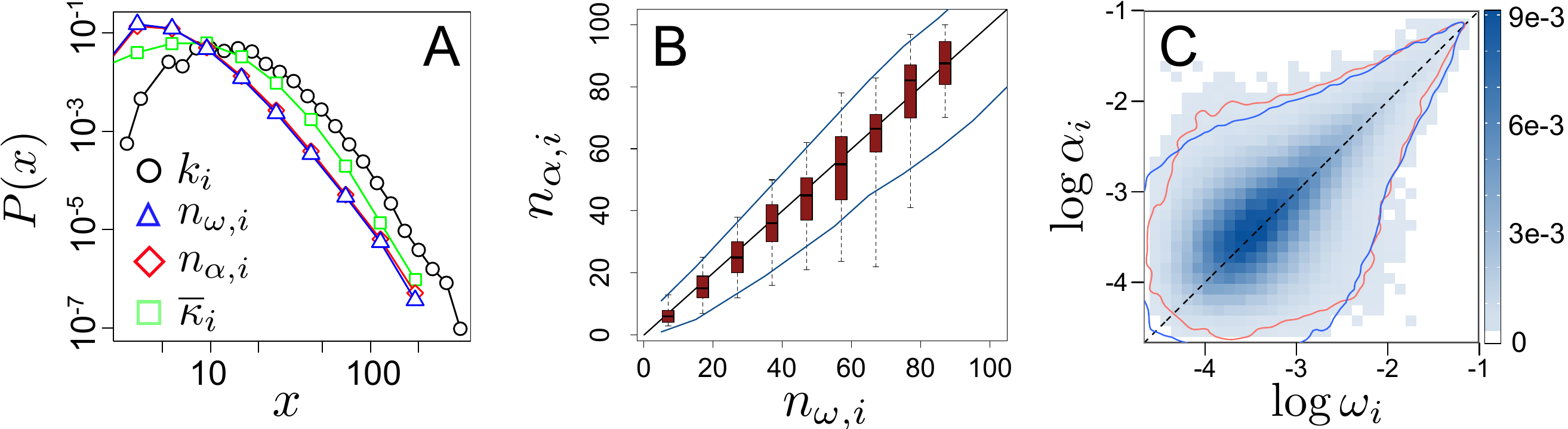}}
\caption{{\bf Characterization of communication capacity and activity} (A) Probability distribution function (pdf) of the aggregated social connectivity $k_i$, number of created ties $n_{\alpha,i}$ and number of deleted ties $n_{\omega,i}$ at $t = T$, compared with the pdf for the average communication capacity $\overline{\kappa}_i$ over the observation window. (B) Relationship
between the number of formed $n_{\alpha,i}$ and decayed $n_{\omega,i}$ ties in the observation window for the users in our database: the results form the PCA  indicate that the 93\% of the variation can be explained by the first component in the (0.70, 0.71) direction, i.e. almost the black line $n_{\alpha,i} = n_{\omega,i}$ in the plot. Furthermore, the box plot shows the 25\% and 75\% percentiles (filled box) and 5\% and 95\% percentiles (whiskers) and the blue curves correspond to the 5\% and 95\% percentiles of the corresponding Poisson null model for our data (see {\em SI Section E}).
(C) Density plot $\rho(\log \omega_i,\log \alpha_i)$ for users with more than 5 ties formed and decayed. Dashed line is the $\alpha_i = \omega_i$ relationship and the curves correspond to the contour lines $\rho = 0.01$ for the density of actual values of rates (red) and the ones obtained in the Poissonian null model (blue, see {\em SI Section E} for further information). 
\label{fig3}}
\end{figure*}

The procedure described above allows us to determine the tie activation and deactivation events for each individual along the observation period of $7$ months (see Fig.~\ref{fig2}). With those events, we build her instantaneous {\em communication capacity} $\kappa_i(t)$, defined as the number of active ties at any given instant $t$. In principle, $\kappa_i(t)$ is very different from $k_i(t)$, the aggregated number of revealed ties up to time $t$, which is usually what is taken as a proxy for social connectivity \cite{blondel}. Because of the bursty nature of interactions, $k_i(t)$ has a fictitious nontrivial time dynamics at the beginning of the observation period which is typically ignored in observations (see {\em SI Section B} for its implications). However, if we aggregate the number of activated (deactivated) ties up to time $t$, denoted by $n_{\alpha,i}(t)$ [$n_{\omega,i}(t)$], we get that at the end of $\Omega$ we have $k_i(T) = \kappa_i(0) + n_{\alpha,i}(T)$. Thus $k_i(T)$ is a combination of the communication {\em capacity} and communication {\em activity} in $\Omega$. In our database we find a large heterogeneity in $n_{\alpha,i}(T)$ and $n_{\omega,i}(T)$ [see Fig.~\ref{fig3}a]: while on average people activate/deactivate about $8$ (reciprocated) ties in a period of $7$ months, $20\%$ of users in our database activate/deactivate more than $15$ ties in that period. Note that on average $n_{\alpha,i}(T)$ and $n_{\omega,i}(T)$ almost equals $k_i(T)/2$, (see Fig.~\ref{fig3}a), which suggests that a large fraction of the revealed aggregated social connectivity $k_i(T)$ is given by newly activated or deactivated connections; similar ratio of activation/deactivation is found in the Facebook database (see SI Section H). Thus, $k_i(T)$ usually overestimates the instantaneous human communication capacity of maintaining active social ties.

The imbalance between the number of activated or deactivated ties measures how communication capacity changes. At the end of the observation period the change is $\kappa_i(T) - \kappa_i(0) = n_{\alpha,i}(T) - n_{i,\omega}(T)$. Interestingly, we find that for most users in our database we get $n_{\alpha,i}(T) \simeq n_{\omega,i}(T)$ (see details in Fig.\ \ref{fig3}b). This means that there is a conservation principle in social communication, where the number of deactivated ties equals the number of activated ties in our observation window $\Omega$ such that the total number of active ties remains almost constant after $T = 7$ months. This conservation of communication capacity not only happens at this particular time scale $T$ but also instantaneously: as seen in Fig.\ \ref{fig2}c for a particular user and in the {\em SI Section D} we find that for around $90$\% of the users tie activation/deactivation happens linearly in time so that $n_{\alpha,i}(t) \simeq \alpha_i t$ and $n_{\omega,i} \simeq \omega_i t$, where $\alpha_i$ and $\omega_i$ are the rates of tie activation/deactivation and $\alpha_i \simeq \omega_i$ (see Fig.\ \ref{fig3}c). These two facts have a remarkable consequence: despite ties are activated/deactivated continually, the communication capacity for each individual remains almost constant throughout the observation period $\kappa_i(t) \simeq  \overline{\kappa}_i$, signaling that people tend to balance the activation/deactivation of ties in such a way that the number of active relationships remains stable over time. The conservation of social capacity is the root of many observations in the literature (see for example \cite{hidalgo,kossinets}) that the distribution of connectivity in social networks seems to be stable in time but the neighbors of a given node change from one time window to another one. Specifically, we find that the average user social persistence ${p}_i$, measured as the fraction of neighbors present at the beginning of the observation window $\Omega$ that remain active until its end, lies around $75\%$. This means that users renew their social circle slowly, in line with studies in off-line social networks \cite{burt}. This value is much larger than what is expected in a model where all ties have the same probability to be activated or deactivated, in which case we obtain $\overline{p'}_i = 50\%$ (see {\em SI Section F}). Our results corroborates that the way in which people activate and deactivate ties from their social network is not random; instead, some existing ties are more probable to be deactivated than others.

\begin{figure*}[t]
\centerline{\includegraphics[width=\textwidth]{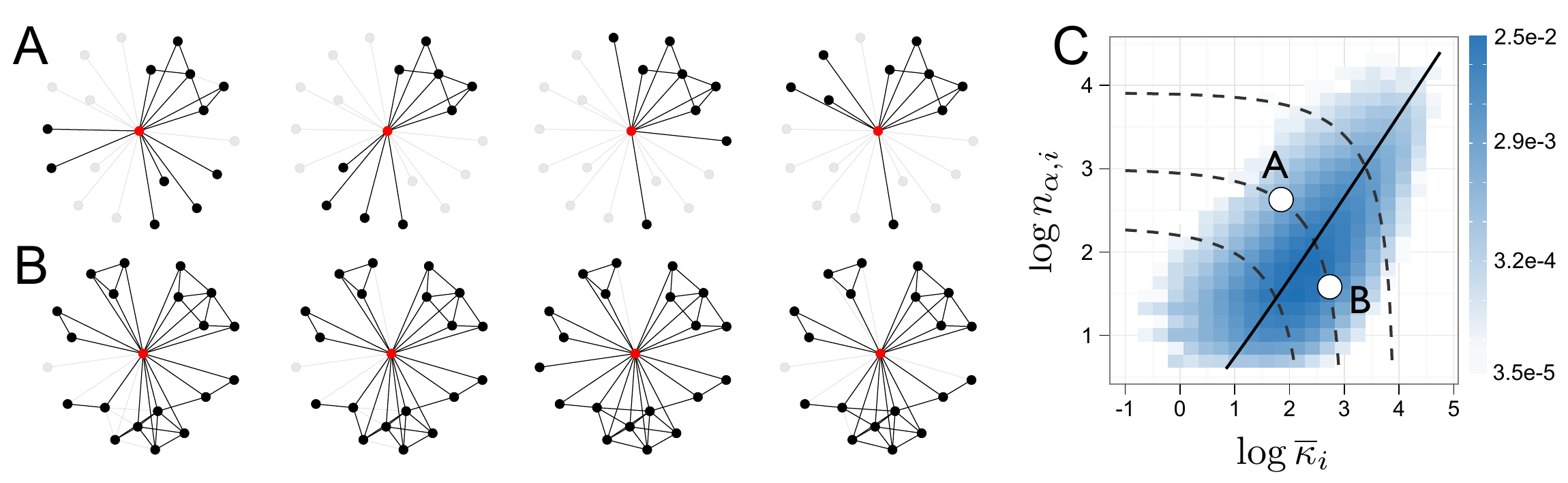}}
\caption{{\bf Variability of communication capacity and activity:} (A) and (B) show different snapshots of the neighborhood of two different individuals (in red) at 4 equally spaced times in the observation time window $t = 52, 105, 158$, and $211$ days. Each black (grey) line corresponds to an active (inactive) tie at that particular instant. (C) Log-density plot of the communication activity $n_{\alpha,i}$ as a function of the communication capacity $\kappa_i$ for each individual in our database. Solid line corresponds to the line $n_{\alpha,i} = 0.75 \overline{\kappa}_i$ obtained through PCA. Dashed curves are the iso-connectivity lines $k_i = \overline{\kappa_i} + n_{\alpha,i}$ for $k_i = 10,20,50$. \label{fig4}}
\end{figure*}

Thus, individual communication can be characterized in terms of his {\it communication capacity} $\overline{\kappa}_i$ and his {\it communication activity} $n_{\alpha,i}$ (or rate $\alpha_i$) in a time window. These two quantities give information about two related although not equivalent features of social communication. While the capacity is a measure of the number of relations that a user manages instantaneously, the activity is instead related to the number of relations a user establishes and at what rate. However, as shown in Fig.~\ref{fig4}, we observe for a large part of the individuals that $n_{\alpha,i} \simeq \beta\,\overline{\kappa_i}$ with $\beta = 0.75$, meaning that the number of created connections tends to be proportional to the communication capacity. This correlation resembles the {\em preferential attachment} process by which tie activation is more probable for more connected individuals. Note however that we find that tie activation is here proportional to a conserved quantity and thus grows linearly in time for $t \gg 1$; and on top of that, there is a corresponding {\em preferential de-attachment} mechanism meaning that individuals with large $\overline{\kappa}_i$ are also more likely to deactivate ties. Although the dependence $n_{\alpha,i} \simeq \beta\,\overline{\kappa_i}$ explains most of the observed behavior (80\% of variance in PCA), there is a still a large variability in our database so that tie evolution cannot be explained solely by $\overline{\kappa}_i$. As shown in Fig.~\ref{fig3}, for a given number of people contacted in the observation period $k_i(T)$ there are many possible combinations of social activity $n_{\alpha,i}$ and capacity $\overline{\kappa}_i$ which yield to the same $k_i(T)$.

\subsection{Lifetime evolution and sex differences}
\begin{figure*}[t]
\centerline{\includegraphics[width=0.8\textwidth]{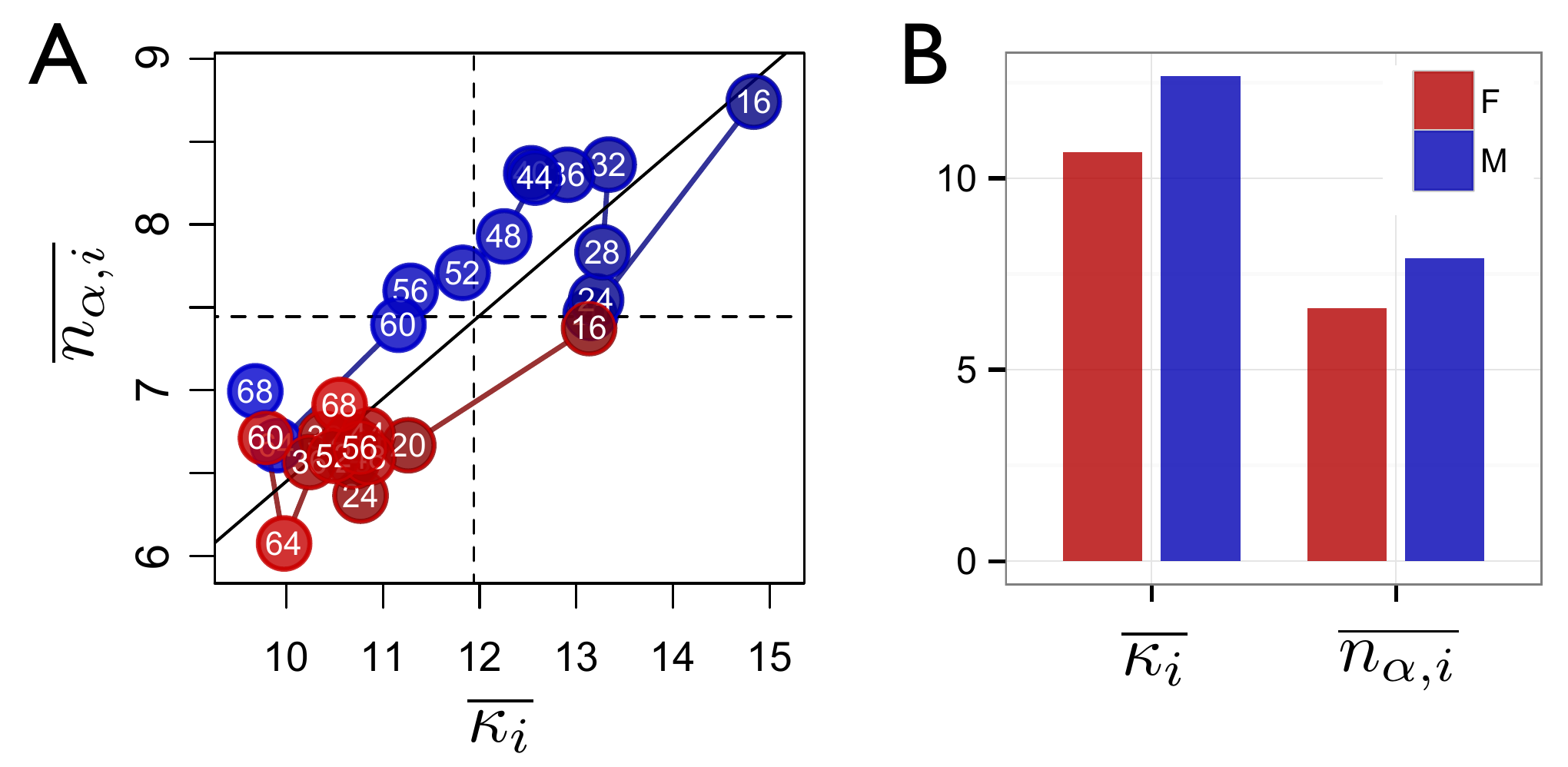}}
\caption{{\bf Sociodemographic dependence of the capacity and activity:} (A)  Average value of the social capacity $\overline{\kappa_i}$ and the activity $\overline{n_{\alpha,i}}$ for groups of users with different age and gender. Dashed lines correspond to the average of $\kappa_i$ and $n_{\alpha,i}$ in the complete database and the solid line is the line $n_{\alpha,i} = \beta \kappa_i$ obtained through the PCA in the complete database. (B) Average values for the activity and capacity of users grouped by gender. \label{fig5}}
\end{figure*}

Although the communication capacity and activity remain mostly stable over the observation time window $\Omega$, they tend to change gradually during the individual life course. Specifically, as shown in Fig.~\ref{fig4}, we observe that as people get older the size of their social circle ($k_i=n_{\alpha,i}+\kappa_i$) decreases. This decrease in both the communication capacity and activity observed in Fig.~\ref{fig5} is in line with previous studies on the lifetime evolution of the cognitive and communication capacity of individuals \cite{tilburg1998,kahn1980,seeman2012}. Specifically, changes in egocentric network size across the individual lifespan are usually associated to both experiencing age-specific life events and social goals \cite{wrzus2012}. Other studies relate the decrease in the social engagement (number of social contacts, interaction activity, frequency of communication) across the individual lifespan, to a decrease in the cognitive capacity~\cite{seeman2012}.
Our decomposition of $k_i$ as a combination of $n_\alpha$ and  $\kappa_i$ allows us to better understand the change in social network size across the individual lifespan and its relation with individual communication strategies.

Although the trend in vital trajectories does not change significantly with the gender of the individual, interesting differences are observed between men and women social strategies (Fig.~\ref{fig4}). First, in line with recent studies using mobile phone records~\cite{frias2010,palchykov2012}, we found that on average women maintain smaller social circles  than men, which seem to happen regardless to their age.  Interestingly, communication activity and capacity have a gradual change over the lifetime of men, with no significant drop before the 60s. On the other hand, women have a clearly marked difference between adolescence ($<16$years) and the rest of their lifetime.

\section{Social strategy}
As we show in Fig.~\ref{fig5}, there are many different combinations of communication capacity $\overline{\kappa}_i$ and activity $n_{\alpha,i}$ which yields to the same number of tie activations/deactivations in the observation window $k_i$. We encode that disparity in the ratio $\gamma_i=n_{\alpha,i}/\overline{\kappa_i}$ which we dub as {\it social strategy} and gives information about the balance between the {\it communication capacity} and the {\it communication activity} for a given node: for $\gamma_i \simeq \beta$ (the average behavior), users have a {\em normal} or {\em balanced} social strategy between their communication capacity and activity. Outside this group we find those users with $\gamma_i \ll \beta$ that activate/deactivate a small number of connections compared to their communication capacity, or users with $\gamma_i \gg \beta$ who have a large communication activity compared to their communication capacity. We refer to these two strategies as {\em social keeping} ($\gamma_i \ll \beta $), meaning that these individuals keep a very stable social circle, and {\em social exploring} ($\gamma_i \gg \beta$), meaning that these individuals activate new ties and deactivate existing ones at a high pace.

In the following we study how such different social strategies relate to topological properties and impact the local and global network dynamics as they operate in the time-scales relevant for viral information diffusion.

\subsection{Relation to topological properties}
We find a significant correspondence between social strategy and individuals' local network topology. As mentioned above, users show on average a $75$\% persistence in their ties in 7 months, where the persistence is measured as the fraction of initial ties that remain active during the whole $\Omega$ (see {\em SI Section F}). However, as shown in the {\em SI Section F} this value rises up to $90\%$ for {\em social keepers} with $\gamma_i < 0.2$ and is only $52\%$ for {\em social explorers} with $\gamma_i > 2$. A similar dependency is found for the (aggregated) clustering coefficient $c_i$: as shown in the {\em SI Section F} for a fixed $k_i$, the clustering coefficient for {\em social keepers} doubles that of {\em social explorers}, meaning that for equal $k_i$ the former have less distinct social contexts or structural diversity \cite{ugander} than the latter. Finally, we find that along with the assortativity of $k_i$ in the social networks we get a large assortativity of social strategies with a Pearson coefficient $\rho(\gamma_i,\gamma_{nn,i}) \sim 0.3$ (see {\em SI Section F} for further details). This means that social explorers/keepers tend to gather. These findings render a dynamical picture of the network with very different evolution rhythms: highly clusterized and almost static areas of social keepers live together with extremely volatile groups of social explorers.

Our analysis of the Facebook communication dataset shows that these patterns also hold for users interacting online (see {\em SI Section H} ).

\subsection{Information diffusion} 
\begin{figure*}[t]
\centerline{\includegraphics[width=0.8\textwidth]{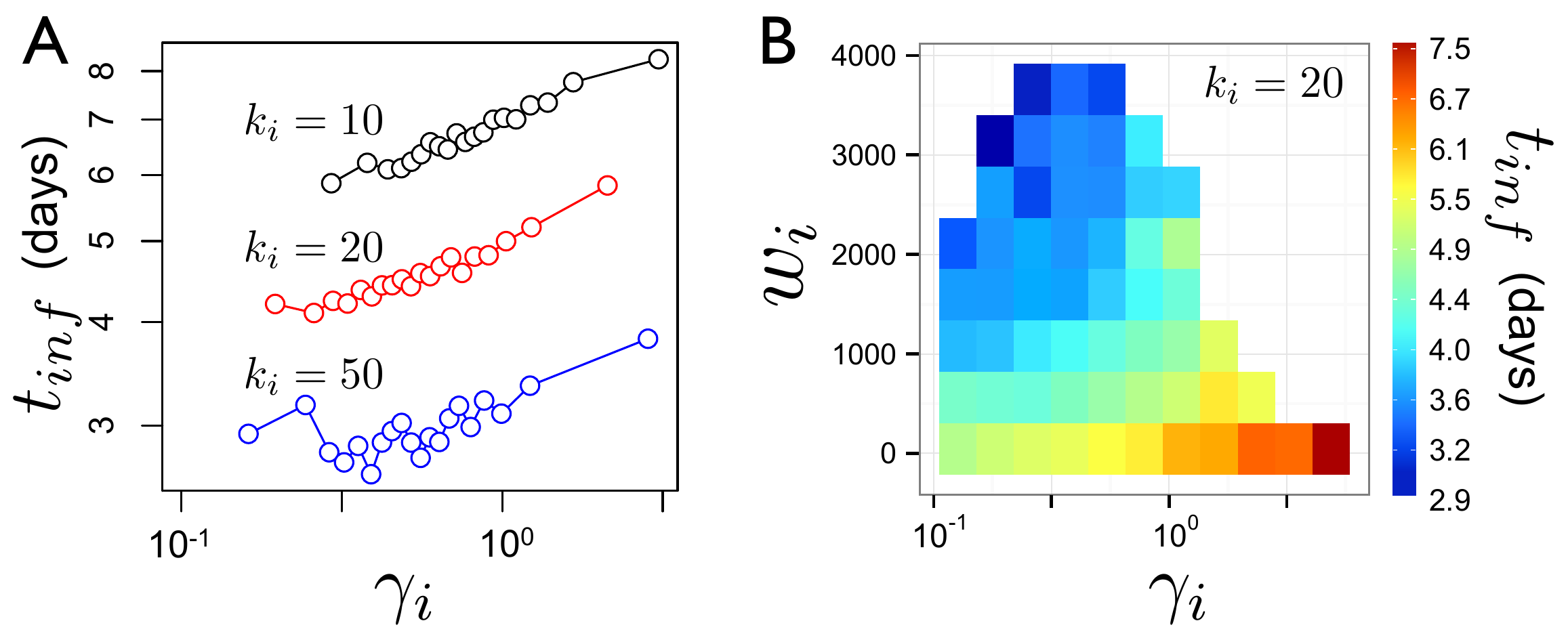}}
\caption{{\bf Infection time and social strategies:} (A) Relation between average infection time and $\gamma_i$ for the different connectivity groups $k_i = 10,20,50$ (the Pearson coefficient between $t_{inf}$ and $\log(\gamma_i)$ is 0.13 with confidence range [0.12,0.14]). (B) For the connectivity group $k_i = 20$ we show the dependence of the average infection time on the total number of exchanged calls $w_i$ and the social strategy $\gamma_i$. \label{fig6}}
\end{figure*}

Finally we investigate whether social strategies have an impact in an individual's capacity to access information being propagated in a network. To address this, we have run the Susceptible-Infected model on the real sequence of CDRs. In a way analogous to  previous works \cite{karsai,miritello}, we start the simulation by infecting a random node at a random time instant and considering all other nodes as susceptible. At each call, if either involved nodes is infected, the susceptible one will be infected too. This maximal spreading process generates a viral cascade which continues until all reachable nodes are in the infected state. We repeat the simulation for $10^4$ randomly chosen seeds. For each individual we then measure the infection time $t_{inf}$ as the time difference between the time at which she received the information and the time at which the corresponding cascade was initiated. Obviously, for a given individual, the infection time decreases with her total connectivity $k_i$ and het total number of communication events $w_i$: the more connections an individual has and the more she interacts, the sooner she receives the information. But when we control for $k_i$ and $w_i$, we observe that on average there is a dependence between how stable the social strategy is and the infection time (Fig.~\ref{fig6}). Interestingly, we observe that social explorers ($\gamma_i > 2$) have a relatively larger time (roughly 2-3 days of difference) to awareness of the information compared to social keepers ($\gamma_i < 0.2$).

We observe that only some combinations of node strength and social strategy are possible. With low to moderate levels of exploration in social strategies ($\gamma < \beta$) it is possible to reach a wide range of node strengths, with a sweet spot in connectivity that allows individuals to lower their time to access information. However, with $\gamma > \beta$ the number of nodes with high strength decreases exponentially: highly exploratory individuals display a very low level of communication events and therefore a very large time to receive information circulating in the network. This result suggest that the information access benefits of diverse ties are outweighted by their short time lifespan, resulting in a net delay in access to information from the individuals activating them.

\section{Discussion}
Our insights can be seen, in essence, as the individual-level dynamical version of the tie-level static results reported by Onnela et al.~\cite{onnela}. The authors analyzed $18$ weeks of mobile phone call records from $7$ million people and showed that, in terms of information diffusion, ties with low cumulative communication time (strength in our context)  are ineffective at information transfer. Our results clarify that these ties are disproportionally generated by social explorers, and that they are mostly activated and deactivated in a short time span. In fact, we find that the average tie weight of each individual $\overline{w}_{ij}$ (measured in terms of average number of exchanged calls per tie) is negatively correlated to the social strategy $\gamma_i$ with a Pearson coefficient $\rho(\log\gamma_i,\log \overline{w}_{ij})\simeq -0.32 \pm 0.01$, indicating that on average weak ties belong mostly to social explorers. Note that these highly time-localized communications differ from the conventional wisdom about weak ties. Typically, in fact, weak ties are seen as bridging connections that span remote parts of the network permanently, since they are considered active over the whole observation period ~\cite{granovetter1973strength,centola2007complex,eagle2010network,centola2010spread}. In our dataset, instead, this happens with low frequency. Although a detailed analysis of what constitutes a weak ties is beyond our scope, we find that of all ties with less than $10$ calls (corresponding to $50\%$ of the whole population of ties), only almost $20\%$ of them remain active during the entire observation window. This is also consistent with the ``Diversity-Bandwidth Tradeoffs'' observed in corporate email communication datasets from two medium sized firms ($107$ people over $10$ months; $214$ over $12$ months). The authors found empirical evidence that people who form ties to disparate parts of the social network at the cost of reducing their band-width of communication can have disadvantaged access to novelty they receive~\cite{sinan2011,sinan2012}. Our simulation results support this result for a large scale social network and connect it to measurable individual strategies. 

Although, as we have seen, the adoption of social strategies does not seem to depend on the magnitude of activity and capacity, we have found them to be assortative. In addition, despite we cannot establish causality with our methodology and observational period, it is an interesting question whether social strategies can be behind the homophily in static topological properties, which has been observed in a wide range of real social networks~\cite{newman2002}.

These findings document an important contrast between possible social dynamics: for almost any given $k_i$ we can find social explorers with that connectivity that navigate the network for new ties and thus have larger structural diversity, as well as social keepers, more conservative individuals who focus attention to their stable social neighborhood. In other words, individuals can exhibit exploratory or stable strategies at multiple scales of connectivity, and these strategies have more important impact in the resulting network properties, ranging from cohesiveness to information diffusion, that the total number of contacts they are able to initiative or receive. This result is important as it provides conclusive evidence for the divergence between the static and dynamic characterizations human interaction. Fine-grained, longitudinal and cross-sectional data as the one presented in this study are then needed to fully understand processes such as navigation, influence and information diffusion as they happen concurrently and possibly entangled to the unfolding of social strategies in time.

\section*{Acknowledgments}
We would like to thank Telef\'onica for providing access to the anonymized data. E.M. and G.M. acknowledge funding from Ministerio de Educaci\'on y Ciencia (Spain) through projects i-Math, FIS2006-01485 (MOSAICO), and FIS2010-22047-C05-04. M.C. acknowledges support from the National Science Foundation under grant 0905645, from DARPA/Lockheed Martin Guard Dog Program PO\# 4100149822, and the Army Research Office under Grant W911NF-11-1-0363.

\newpage

\appendix

\ \ \ 
\newpage

\twocolumn[\begin{@twocolumnfalse}

\noindent {\bf \Large Supporting Information for}\\

\noindent{\LARGE \em Limited communication capacity unveils dynamical strategies for human interaction}

\vspace{0.1cm}

\noindent
{Giovanna Miritello, Rub\'en Lara, Manuel Cebri\'an, and Esteban Moro}

\vspace{0.2cm}
\hrule
\vspace{0.4cm}

\end{@twocolumnfalse}]

\singlespace
\section*{A. Preparing and Sampling the Data}\label{SIs1}
The data used in this study has been obtained from the Call Detail Records database of a unique mobile phone operator in a single country. We focused exclusively on voice calls records, filtering out short text messages, multimedia messages and operator calls. Each subscription is anonymized such that it is not possible to recover personal information of the users. We filtered out all the incoming or outgoing calls that involve other operators due to the partial access we have to the activity of other providers. To avoid business-like subscriptions, which usually appear as users with a huge number of connections and calls never returned, we only retain ties which are reciprocated, which leads to the removal of about the 50\% of the total links in our database. This restriction also eliminates calls to wrong numbers, telemarketing-type calls, customer service lines, etc. Within this approach, we neglect the directionality of links and consider a call from $i$ to $j$ equivalent to a call from $j$ to $i$ \cite{onnela}. The resulting mobile graph contains the communication of about $20 \times 10^6$ and users over a period of 19 months from February 2009 to August 2010.

To disentangle the dynamics of ties creation/removal from their call activity, we split the 19-months period into 3 subintervals (Feb09 - Jul09, Aug09 - Feb10, Mar10 - Aug10), (see {\em Main text Fig.\ \ref{fig1}}). We have only considered the evolution of the ties and nodes that show any activity in the 7 months observation window $\Omega$. The resulting graph in $\Omega$ contains $16 \times 10^6$ individuals and $130\times 10^6$ ties. The intervals before and after are used to assess respectively whether the ties exist from before and/or persist after $\Omega$. Fig.\ \ref{fig1} shows the different situations that can occur for a given tie. In particular, in our database, the 12.5\% of links belongs to the category (a), the 14.5\% to (b), the 22.2\% to (c) and the 47.3\% to (d), while only the 3.5\% of the links, which belong to category (e), will be missed in our analysis.

Since we are interested only in tie dynamics between individuals, we have to take into account the problem of subscription and churn of users in our database. For example, subscription of a new user and its communication with other users in our database results into formation of many new ties for the new subscriber. The same would happen for the decay of ties of a subscribe that churns from the company. To mitigate this problem, we only keep active users in our data set: in particular, we only consider those users who are involved (as calling or as called party) at least in one communication event in each of the three subintervals in the 19 months and also if they are present in the database at least one month before $\Omega$ and are still active one month after $\Omega$. This latter filter prevents spurious effects in the analysis of tie dynamics just because individuals subscribe/unsubscribe just before/after $\Omega$; for example, we could have observed an apparent rapid growth of their social network at the beginning of the observation window or a fast dissolution at its end \cite{gaito}. This results in the removal of about the 17\% of nodes and the 37\% of reciprocated links within $\Omega$.

In our database, we also have information on the age and gender of users of a random chosen fraction (40\%) of them . We found that the minimum and maximum values of age are respectively 0 and 97. However, we only keep users whose age lies between 16 and 70 years old in order to yield a more reliable dataset. This filtering led to the removal of the 0.5\% of users which demographic data.

\section*{B. Entanglement between bursty activity and tie dynamics}\label{SIsection2}
As stressed in the main text, one of the most challenging problems in the study of the dynamics of tie creation and removal is to identify whether a tie is actually a new/old connection. Although in most social networks there are specific events for the formation of new "friends" (or followers) or the corresponding "unfriending" events, due to the cheap cost of maintaing those connections most of those ties are abandoned and thus activity between individuals is the only way to asses the existence or not of that relationship. 

\begin{figure*}
\begin{center}
\includegraphics[width=0.75\textwidth]{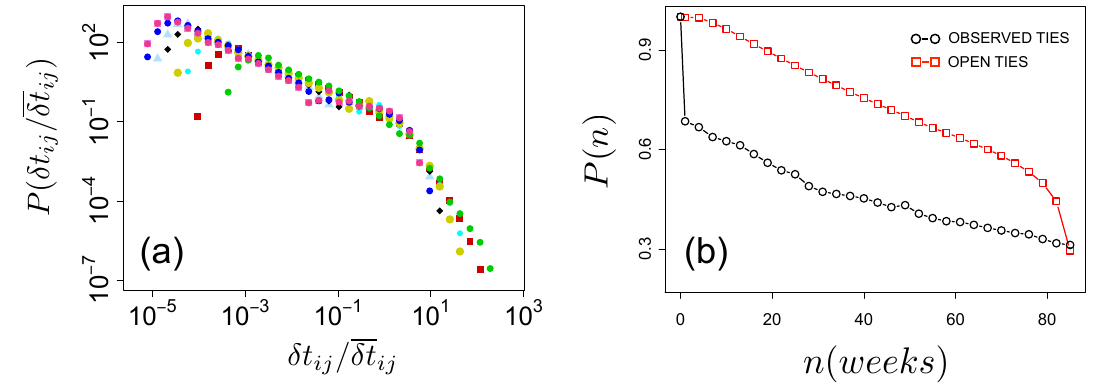}
\end{center}
\caption{(a) Rescaled inter-event time distribution for groups of edges with different average inter-event time $\overline{\delta t}_{ij}$. Each curve is rescaled by the value of $\overline{\delta t}_{ij}$ of the correspondent bin. (b) Weekly persistence  $p(n)$ of ties observed in the first week of our database as a function of the number of weeks $n$: while persistence drops to 70\% after one month if ties are required to have activity at a given week $n$, it is still around 70\% for one year if we consider open ties at that week, i.e. ties which where observed in the first week. \label{fig1SI}}
\end{figure*}

However, human activity is bursty, meaning that there are large periods of inactivity followed by bursts of activity \cite{barabasiburst}. This means that within a particular tie $i \leftrightarrow j$ the time between consecutive communication events $\delta t_{ij}$ is heavy-tailed distributed. In our database we find that this is indeed the case and in line with \cite{karsai,miritello} we find that there is a universal law for the distribution of inter-event times (see Fig.~\ref{fig1SI}). In particular, we find that for a particular tie $P(\delta t_{ij}) = {\cal P}(\delta t_{ij} / \overline{\delta t}_{ij})$ where ${\cal P}(x)$ is a heavy tailed universal function. Since bursty behavior seems to be universal in human activity \cite{barabasiburst}, it has a deep impact in the understanding of tie dynamics and translate in a ubiquitous problem in the empirical observation of social networks: if the observation window is very short we might miss most of the ties since there is no communication in that period of time. But on the other hand, since the inter-event time distribution is heavy-tailed we might have to go to large observation windows to recover most of the ties. For example in our database we find that the average inter-event time is $\langle \delta t_{ij} \rangle = 14$ days with a standard deviation $\sigma = 18$ days, which means that the observation period must be larger than a 2-3 months only to observe (at least once) most of the ties in the social network. In our case, $\Omega$ extends over $7$ months and using the previous/next 6 months intervals we calculate that $3\%$ ties did not show activity in $\Omega$ and then could have been missed if only data within $\Omega$ was present.  

Although the impact of burstiness in the observation of ties is important, it becomes critical for the problem of tie formation/decay since it is not only necessary to observe the tie but to asses its termination or formation. Thus, we need to increase substantially the observation window to identify whether the link has been formed and or decayed in our database. Short observation windows can lead to spurious effects: a tie that is present in one time window might (with large probability) do not show activity in the next time window due to a large inter-event time and thus we might incorrectly identify that event as decay of the relationship. This might be the origin of the large (30-40\%) decay in persistence observed in the literature \cite{hidalgo,raeder} (and reproduced in our database, see Fig.~\ref{fig1SI}(b)), since the observation windows were very short (1 month). The large probability of having a inter-event time of one month in human communication leads to the erroneous impression that 40\% of the links are created/decayed in one month period and that the networks are highly volatile, since correlation between the network structure at different observation windows is very low. 

To cure those problems in our paper we propose a different method to asses whether a tie formed/decayed in the observation window $\Omega$. The method is based on the {\it observation} of tie activity in a time window before/after $\Omega$: if tie activity is observed in the 6 months before $\Omega$ then it is considered an old tie [cases (a) and (d) in {\em Main Text Fig.\ \ref{fig1}}]; on the other hand, if activity is observed in the 6 months after $\Omega$ we will assume that the tie persists [cases (b) and (d) in {\em Main Text Fig.\ \ref{fig1}}]. In any other case, we will consider that the tie is formed and/or decay in $\Omega$ [cases (a), (b) and (c) in {\em Main Text Fig.\ \ref{fig1}}]. Of course, it is possible that even if there is no communication before/after the observation window, the tie is still active after/before our database. This would require that the tie has an inter-event time $\delta t_{ij}$ bigger than 7 months, i.e. case (e) in {\em Main Text Fig.\ 1}. However, in our database, only 3.5\% of the links have such a long inter-event time which validates the accuracy of our definition of tie decay/formation. 

On the other hand in our study a tie is considered to be opened between  its formation and decay events (if they happen in $\Omega$ at all). This assumption is based on the idea that an interaction which has been observed in the past and will be observed in the future might exist at a given instant even if there is no communication by mobile phone between at that particular instant. Furthermore, our observation window is short enough to neglect safely possible formation and decays of the relationship within $\Omega$. Our definition of relationship mitigates the excessive volatility of the social network when tie is considered only when interaction is observed at a given instant. For example, the persistence of open links is higher (70\% in one year) than observed links (40\% in one year) in line with off-line studies \cite{burt}.  It also resembles different situations in which, although a strong relationships might exist off-line, very few calls are exchanged in time. 

Finally, understanding this difference between open and observed relationships is crucial to unveil the real dynamics of social networks because it can induce also spurious effects in the observations: within a given observation window, the (revealed) aggregated connectivity $k_i(t)$ seems to grow non-trivially as a function of time (see {\em Main Text Fig.\ \ref{fig2}}) within $\Omega$. Actually, it could even be fitted to a power law $k_i(t) \sim t^{\gamma}$ with $\gamma \simeq 1/2$ for small $t$. It is interesting to see that the functional form and exponent do fit those found in models of network growth \cite{newmanSIAM}. But it is easy to see that this effect is (mostly) due the fact due the fact that different links have very heterogeneous number of communication events $w_{ij}$ and within a given tie events are very bursty. Specifically, the apparent growth of $k_i(t)$ for short times is mainly due to the possibly large and highly heterogeneous time to the first event event within ties. 

To understand that, suppose that a given tie is present before and after the observation window $\Omega$ and that the distribution of inter-event times within that tie is given by $P(\delta t_{ij})$. Assuming that the initial time of the observation window is random, the time to the first observation of the link is given by the waiting time equation in renewal processes \cite{breuer} 
\begin{equation}
P(\tau_{ij}) = \frac{1}{\overline{\delta t_{ij}}} \int_{\tau_{ij}}^\infty P(\delta t_{ij}) d\delta t_{ij}
\end{equation}
Thus, depending on the properties of $P(\delta t_{ij})$ we could have a very large observation time ($\tau_{ij}$) for the link. As shown in Fig.~\ref{fig1}(a) the pdf for inter-event times depends mostly on the average inter-event time $\overline{\delta t}_{ij}$, i.e. $P(\delta t_{ij}) = {\cal P} (\delta t_{ij} / \overline{\delta t}_{ij})$ where ${\cal P}(x)$ is a universal function. Thus, for a given $\overline{\delta t}_{ij}$ we could rewrite the previous expression as
\begin{equation}
P(\tau_{ij} | \overline{\delta t}_{ij}) = \frac{1}{\overline{\tau_{ij}}} \int_{\tau_{ij}}^\infty {\cal P}(\delta t_{ij} / \overline{\delta t}_{ij}) d\delta t_{ij}
\end{equation}

However, ties are very heterogeneous in the sense that they have very different $\overline{\delta t}_{ij}$. Or equivalently, they have very different weights $w_{ij} = T/\overline{\delta t}_{ij}$  \cite{onnela}. Suppose that $\Pi(\overline{\delta t}_{ij})$ is the distribution of average inter-event times across links and that each user chooses her tie activities from that distribution. We assume also that no tie is form/destroy during the observation time. Then the probability to observe one of her links at time $\tau$ is given by:
\begin{equation}
P(\tau) = \int d \overline{\delta t}_{ij} \Pi(\overline{\delta t}_{ij}) P(\tau | \overline{\delta t}_{ij})
\end{equation}
Thus, the growing function of the observed connectivity as a function of time is given by the ccf of $P(\tau)$.
\begin{equation}\label{growthki}
k_i(t) = k_i(T) \int_0^t P(\tau) d\tau 
\end{equation}
where $k_i(T)$ is the total connectivity of node $i$ in the observation window $\Omega$. Note that since ${\cal P}(x)$ and $\Pi(\overline{\delta t}_{ij})$ are heavy tailed, then $P(\tau)$ is heavy tailed too and thus the $k_i(t)$ can show an apparent non-trivial time dependence even if all links are open during $\Omega$. Expression (\ref{growthki}) shows that one should be careful to consider the observed aggregate connectivity $k_i(t)$ as a proxy for social connectivity at any time $t$, since it is profoundly affected by the bursty and heterogeneous activity of human behavior encoded in $P(\tau)$. Note to mention the effect of tie formation/destruction which is not included in (\ref{growthki}).

\begin{figure}
\begin{center}
\includegraphics[width=0.45\textwidth]{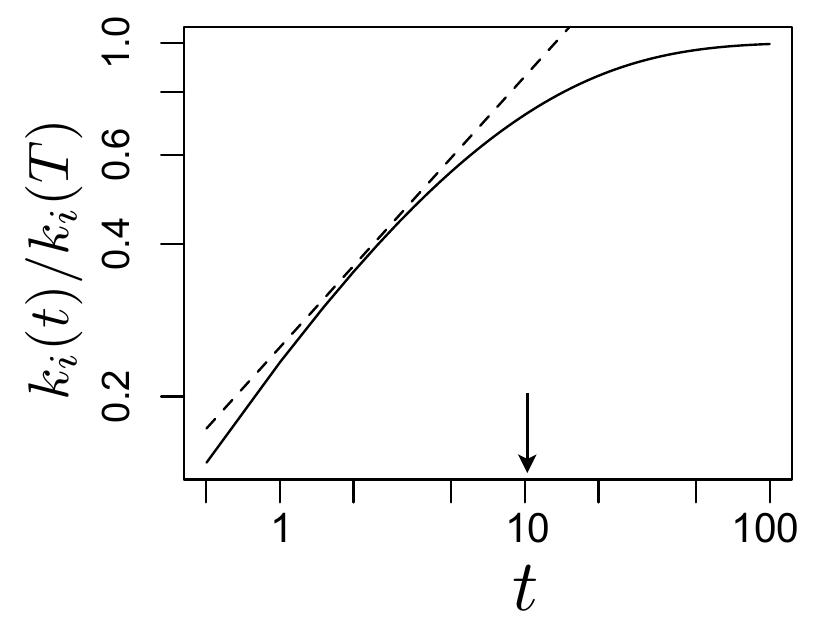}
\end{center}
\caption{Apparent growth in the connectivity given by equation (\ref{growthki}) as a function of time for an exponential distributions of average inter-event time with $a = 10$ days (marked by the arrow). Dashed line is the fit to a power-law growth for the initial growth (up to 5 days) that yields $k_i(t) \sim t^\gamma$ with $\gamma = 0.53\pm0.02$.\label{fig2SI}}
\end{figure}

Strikingly, an apparent $k_i(t) \sim t^{\gamma}$ growth can be observed even in the case in which both tie activity and weights are severely bounded: if we assume that the distribution of inter-event times is given by the exponential pdf $P(\delta t | \overline{\delta t}) = e^{-\delta t / \overline{\delta t}}/\overline{\delta t}$ and also that the pdf for the average inter-event time is an exponential $\Pi(\overline{\delta t}) = e^{-\overline{\delta t}/a}/a$ we get exactly from Eq.~(\ref{growthki}) that 
\begin{equation}
k_i(t) = k_i(T) \left\{ 1 - 2 \sqrt{\frac{t}{a}} K_1 \left( 2 \sqrt{\frac{t}{a}} \right) \right\},\qquad 0 \leq t \leq T
\end{equation}
where $K_1(x)$ is the Modified Bessel Function of the second kind \cite{abramowitz}. Thus, for a single user the number of observed ties grows in a non trivial way as a function of time even for this homogeneous (both in the events and in the links properties) case, a behavior which extends further from $t=a$, the average $\overline \delta t$ (see Fig.~\ref{fig2SI}).
This result for a single user based on the universal bursty and heterogeneous activity in ties, together with the large heterogeneity found in social connectivity (which is related to $k_i(T)$) could explain the apparent non-trivial growth of the aggregate $k_i(t)$ observed in social networks \cite{blondel} and highlights the importance of taking into consideration the heterogeneity of activity of humans to define properly the way we measure and observe their social networks. Finally these results emphasize the goodness of our method to detect open ties, since in this simple example all ties are open at any time and then $\kappa_i(t)$ is constant throughout the observation window $\Omega$.

\begin{figure*}
\begin{center}
\includegraphics[width=0.8\textwidth]{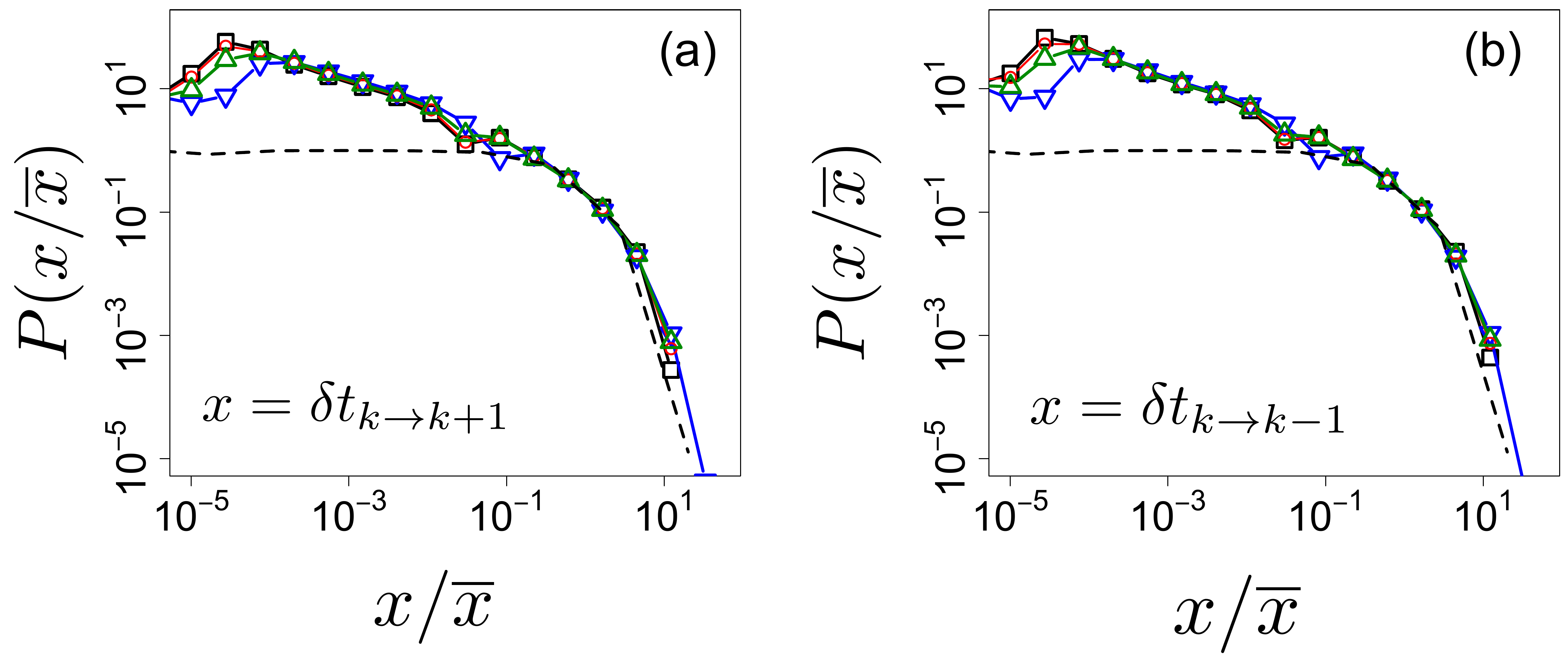}
\end{center}
\caption{(Rescaled) Distribution of the time gap between edge creation (a) and edge removal (b) for groups of nodes with different  activity rate $\alpha_i$, where groups have been obtained according to the quartiles of $\alpha_i$ for the whole population. \label{fig3SI}}
\end{figure*}

\section*{C. Bursty dynamics of tie activation or deactivation}\label{SIsection3}
We observe that most people form and destroy edges almost constantly in time (see {\em Main Text Fig.\ \ref{fig3}}). However, despite the linear growth of the number of added and removed connections, the distribution of the time gap between creation/removing of ties is not Poissonian (Fig.\ref{fig3SI}), which is in line with recent results \cite{karsaiskype}. Fig.~\ref{fig3} (a) and (b) show respectively the pdf of the time it takes for the node $i$ with degree $k_i$ to create one more connection ($\delta t_{k,k+1}$) and to loose one connection $\delta t_{k,k-1}$. Specifically, we 
divide the whole population of users in four groups depending on their value of $\alpha_i$ and plot the distribution for each group.  Despite the exponential cut-off, the results indicate some bursty patterns of activity for sort times. In addition all distributions collapse into a single curve suggesting that a universal form for the burstiness in the tie activation/deactivation.

\section*{D. Linear growth of tie activation or deactivation}\label{SIsection4}
\begin{figure}
\begin{center}
\includegraphics[width=0.5\textwidth]{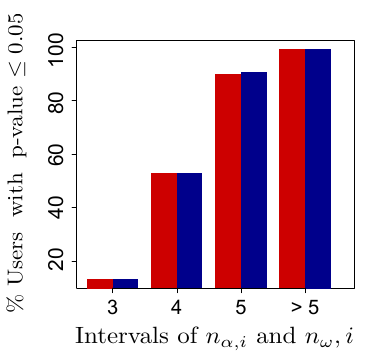}
\end{center}
\caption{Ratio of the number users for who a linear fit to the $n_{\alpha,i}(t) \sim \alpha_i t$ (red) and $n_{\omega,i}(t) \sim \omega_i t$ (blue) time series has a $p$-value smaller than $0.05$ for the F-test. Different columns refer to different groups of users according to their total number of activated/deactivated ties in the observation period $\Omega$.\label{fig8SI}}
\end{figure}

Although tie activation/deactivation events do not happen homogeneously in time, the strong cut off in the bursty inter-event time found in the previous section suggests that there exists a typical time scale in which those events happen and thus, for a larger enough observation time, we should expect linear growth for the accumulated number of events $n_{\alpha,i}(t)$ and  $n_{\omega,i}(t)$. Indeed, by taking these time series and fitting them to linear models we get the rates $\alpha_i$ and $\omega_i$ explained in the main text. The statistical significance of the fit of those to each individual dynamics is shown in Fig.~\ref{fig8SI} where we can see that the linear fit is statistically significant for a majority of users with $n_{\alpha,i}(T) = 5$ and for most users with $n_{\alpha,i}(T) > 5$ (same results for $n_{\omega,i}$). On the other hand, for those selected individuals for which $p$-value < 0.05 the goodness of fit is on average $R^2 \simeq 0.91$ with 93\% of them with $R^2 > 0.8$. Thus, the results presented in the following section and in the main text for $\alpha_i$ and $\omega_i$ are only for those with $n_{\alpha,i}(T) \geq 5$ (same for $n_{\omega,i}(T)$) for which the goodness of fit is around $R^2 \simeq 0.91$ and the percentage of those with a $p$-value smaller than 0.05 is around $100\%$. They amount up to 75\% of the total number of users.

\section*{E. Statistical evidence for the conservation of social capacity}\label{SIsection5}
One of the key findings in our study is the fact that for a given individual $i$ the rate at which ties are formed $\alpha_i$ equals that at which ties decay $\omega_i$. This implies that social capacity, i.e. the number of open connections at a given instant is more or less constant in time. In this section we describe the analysis performed to reach this statement and the null model used to asses the statistical significance of $\alpha_i \simeq \omega_i$. The basic problem is the fact that for most of our users in the database, the number of events $n_{\alpha,i}$ and $n_{\omega,i}$ is very small and then we get large differences between the values of $\alpha_i$ and $\omega_i$ obtained. 

We will test that our results are comparable statistically to a null model in which ties are formed and destroyed in the observation window $\Omega$ according to two different realizations of a Poisson process with the same rate $\alpha = \omega$. The choice of Poisson process as the renewal process that describes the formation and decay process is supported by the bounded probability distribution for the inter-event times between formation/decay of events seen in previous section. Of course this is an approximation, because there is a large probability of bursts of formation/decay events than predicted by the exponential distribution of the Poisson process. The approximation works better for large times or number of events, since in that limit the strong decay of the inter-event time distribution for large values makes the process to converge to the behavior of a Poisson process very quickly by means of the Central Limit Theorem \cite{renewal}.

Since there is a large heterogeneity of social activity in our database we take as input for our null model the actual values of $n_{\omega,i}$ to incorporate that heterogeneity in our null model. We have also done simulations taking $n_{\alpha,i}$ and the results are the same. Thus, our Monte Carlo simulations of the null model are as follows: for every individual $i$ we take $\lambda_i = n_{\omega,i}/212$ as the rate for tie formation and decay of ties per day and simulate two Poisson processes in the observation window $\Omega$ with the same rate, one for the formation of ties and the other for the decay of ties. We then calculate the times series of the aggregate number of events $\hat n_{\alpha,i}(t)$ and $\hat n_{\omega,i}(t)$ and fit them to linear models to obtained the simulated $\hat \alpha_i$ and $\hat \omega_i$. In line with the results of previous section, we only consider for the fit those simulations for which the $\hat n_{\alpha,i}(T) \geq 5$ and $\hat n_{\omega,i}(T) \geq 5$ in the fit.

As shown in the caption of {\em Main Text Fig.\ \ref{fig3}b} (see details there), the observed values for $n_{\alpha,i}(T)$ and $n_{\omega,i}(T)$ in our database can be well explained by our simulations, suggesting that our model works well at that particular time scale. We also find a good agreement between the measured values of $\alpha_i$ and $\omega_i$ and the results of our null model as shown in {\em Main Text Fig.\ \ref{fig3}c}, although there a small amount of outliers that cannot be explained by our model.

\section*{F. Measuring neighborhood persistence}\label{SIsection6}
We measured the persistence $p_i$ of a user $i$ as the fraction of his neighbors present at the beginning of the observation period $\Omega$ that are maintained until the end of $\Omega$. Specifically, $p_i=({\cal E}_i(0)\cap {\cal E}_i(T))/{\cal E}_i(0)$, where ${\cal E}_i(0)$ and ${\cal E}_i(T)$ are respectively the set of ties that user $i$ has at time 0 and time $T$ (see {\em Main Text Fig.\ \ref{fig1}}). Once measured $p_i$ for all users in our dataset, we find that the average persistence $\overline{p}_i$ is 0.75. As discussed in the main text, this suggests that although in a given time period users activate and deactivate many connections (on average half of their social connectivity), after a period of 7 months they maintain on average the 75\% of their initial social network. 

We also mentioned that this value is much larger than the one obtained in a model in which each tie is activated and deactivated with the same probability, suggesting that as expected individuals do not establish or remove social connections randomly. To address the latter, we simulated the following process: for a given user we preserve {\it(i)} all the properties of his measured social strategy ($k_i$, $\kappa_i$, $n_{\alpha,i}$, $n_{\omega,i}$) and {\it(ii)} the real sequence of both his tie activation and deactivation time instants.
Thus, following the order of such sequences, at each activation (deactivation) time we allow the user to add (remove) one of his neighbors randomly chosen among all his neighbors.
Note that in the random model we maintain all the properties of the individual social network and strategy and the only thing that we destroy is the selection of neighbors added and/or removed within the observation period $\Omega$.
We then repeat this process for all users in our dataset and for each of them we measure the new network persistence $p'_i$. As discussed in the main text, we found that $\overline{p'_i}=50\%$, against the $\overline{p}_i=75\%$ measured for the real case, which suggests that the way in which people activate, maintain and deactivate social relationships is, as expected, not random and some ties are more probable to be destroyed than others.

\begin{figure*}[t]
\begin{center}
\includegraphics[width=0.75\textwidth]{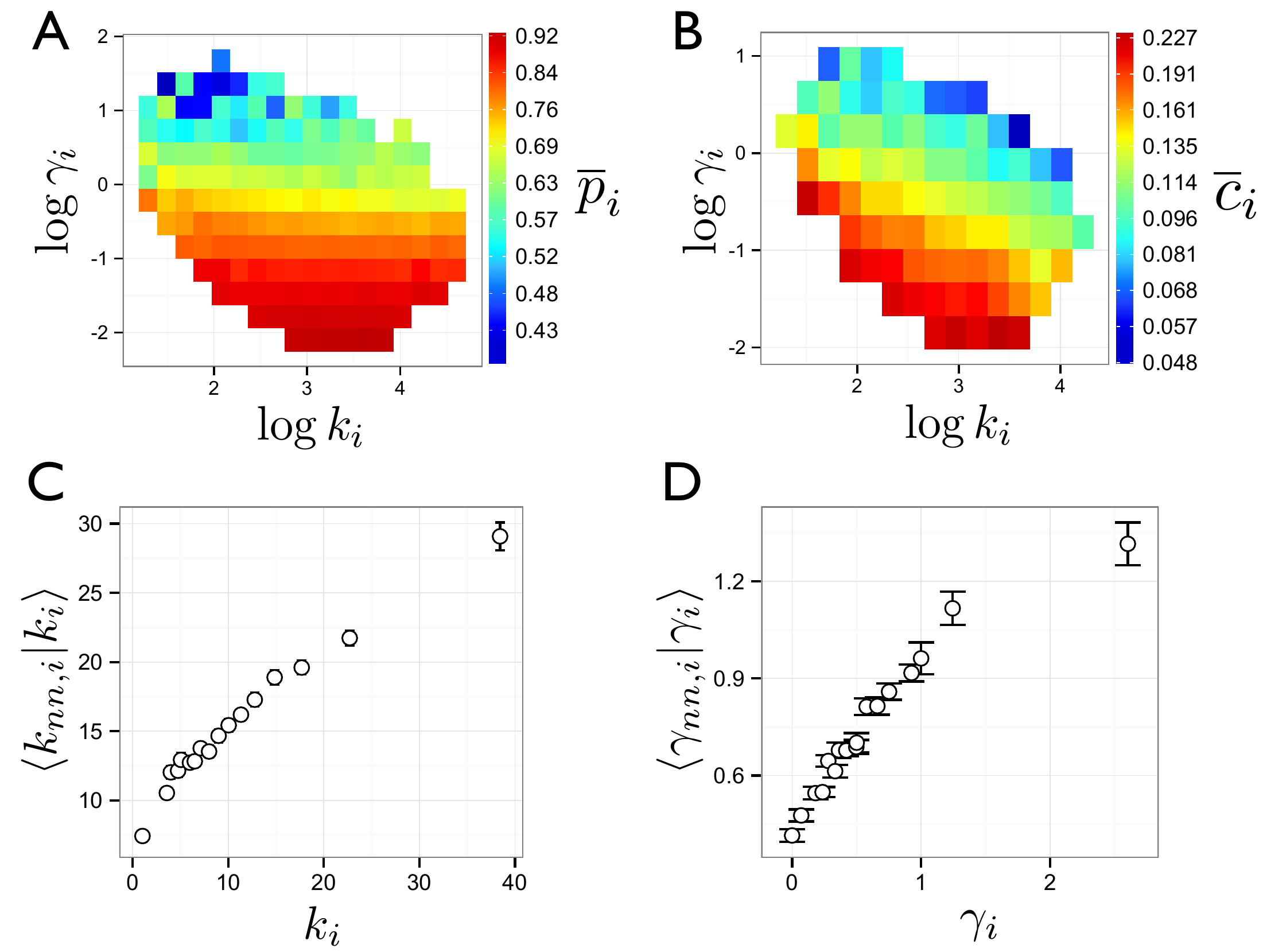}
\end{center}
\caption{Relation of the social strategy with topological properties: dependence of the average persistence of ties (A) and aggregated clustering (B) as a function of the total connectivity $k_i$ and social strategy $\gamma_i$. (C) Average value of next neighbor connectivity $k_{nn,i}$ of a node as a function of its own connectivity $k_i$. The Pearson correlation coefficient between the two quantities is $\rho(k_i,k_{nn,i})=0.342$ with a confidence range of [0.278,0.316] (D) Average value of the parameter $\gamma_{nn,i}$ for the neighbors of an individual as a function of her own value of $\gamma_i$, $\rho(\gamma_i,\gamma_{nn,i})=0.412$ with confidence interval [0.394,0.429]. A clear growth can be seen in both cases, indicating a strong assortativity. \label{fig4SI}}
\end{figure*}

\section*{G. Relation of the social strategy with topological properties}\label{SIsection7}
We find a significant dependence between the social strategy for an individual (encoded through the parameter $\gamma_i$) and the topological properties around that individual. Specifically, figure \ref{fig4SI} shows how the persistence defined in the previous section depends heavily on $\gamma_i$ but shows a large independence with the total connectivity of individuals in the period of observation. Specifically social keepers (those with $\gamma < 0.2$) do show a large persistence in their social neighborhood (even up to 90\%), while social explorers ($\gamma > 2$) only keep a small fraction of their initial ties at the end of the 7 months period, even down to 40\%. On the other hand, the aggregated clustering coefficient also depends on the social strategy: social keepers tend to have more clustered neighborhoods than social explorers. Specifically, we find that $\overline{c}_i$ can be up to $0.22$ for social keepers, while it decreases to $0.05$ for social explorers. Note that in the case of the clustering coefficient we also observe that it decreases with increasing average connectivity, a effect well known in social networks \cite{newmanSIAM}: $c(k_i)$ is typically a decreasing function with $k_i$ reflecting the fact that for largely connected people it is increasingly more difficult to have a moderate clustering. However, in Fig.~\ref{fig4} we see that the clustering not only depends on the connectivity, but also on the social strategy. Since both factors have opposite effect on clustering we find, for example, that social keepers with large connectivity might have the same clustering as social explorers with small connectivity. Thus the aggregated clustering found in social networks is a function of both connectivity and social strategy, suggesting that its value is determined dynamically by the tie formation/destruction processes around a given individual.

Finally, in our database we observe that social connectivity is assortative, in line with other studies \cite{newmanSIAM}. More interestingly, we find that also social strategies of communication are assortative, as it is shown in Fig.~\ref{fig4}.
As mentioned in the main text, this result indicates that people that establish and remove many connections from their network at a high rate (social explorers) are more likely to interact with people that also change their network quickly.
Analogously, those individuals that maintain a more stable social network (social keepers) also interact with people with the same strategy. As a consequence, the large volatility observed in the neighborhood of social explorers also extends to large proportions of the network around them and the same applies for social keepers. The global network thus consists of almost static zones of social keepers and high volatile clusters of social explorers that, as discussed in the main text, also have important implications in terms of information diffusion.

\section*{H. Facebook data set}\label{SIsection8}
\begin{figure}
\begin{center}
\includegraphics[width=0.45\textwidth]{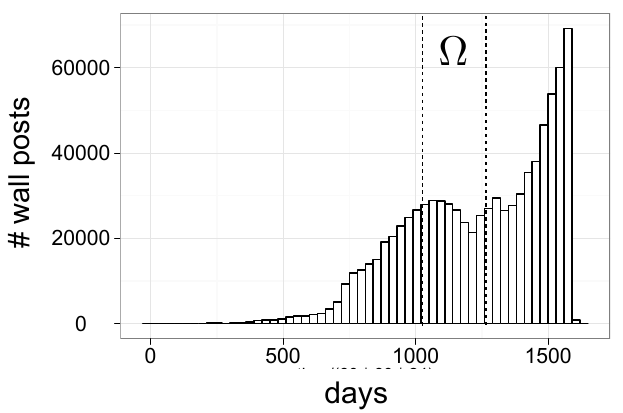}
\end{center}
\caption{Activity in the Facebook database. Number of communications through the wall in our database for periods of 30 days. Dashed lines show the limits of the observation time window $\Omega$.\label{fig5SI}}
\end{figure}

We have also analyzed other communication data to test our results. In particular, we have studied the 90,269 users of the New Orleans Network crawled during by Viswanath {\em et al.} \cite{facebook}. The data consists of communication events between users through Facebook wall from September 26th, 2006 to January 22nd, 2009. Contrary to the mobile phone data, the Facebook data is not steady in time, since the database extends over the early days of Facebook growth and thus it shows a growth in the activity over years, which translates in more wall posts and also more users as a function of time (see Fig.~\ref{fig5SI}).  

To minimize this effect we have chosen only communication events between users that did show any activity in the observation window $\Omega$ (the time interval between 1000 and 1212 days in the database) and also which were present 20 days before and after $\Omega$. We do not consider the links to be reciprocated in order to have more data accessible for our analysis. With this filter our database contains $125 \times 10^3$ communication events of $\sim 10^4$ users and $69 \times 10^3$ ties. 
On average, users interact with $\langle k_i(T) \rangle = 6.15$ users in 7 months and the social activity is $\langle n_{\alpha,i}(T) \rangle = 3.01$, $\langle n_{\omega,i}(T)\rangle = 3.02$ ties formed and decayed respectively. Our results are very similar to the ones observed for mobile phone data, namely that social activity is roughly half of the social connectivity in 7 months. However, users show a lower level of wall activity: for example, 40\% of the users are involved in less than 10 communication events through the wall in seven months (while in the mobile phone data the average number of calls exchanged per user was $\sim 700$ in seven months). Thus, to determine the social dynamical strategies in Facebook data we concentrate on those users that show a moderate level of communication, i.e. those that have more than 10 events in the 7 months of $\Omega$. For those users in our database we find that $n_{\alpha,i}(T) \simeq n_{\omega,i}(T)$ and $\alpha_i \simeq \omega_i$, signaling that users in Facebook tend also to conserve the number of open connections $\kappa_i(t)$ in time (see Fig.~\ref{fig6} (a) and (b)) . On average we find that $\langle \kappa_i(t)\rangle = 3.23$. Finally, as in the mobile phone data we find also a relationship between the capacity and the activity of users: in particular, 81\% of the variance can be explain by the relationship $n_{\alpha,i} = 1.04 \overline {\kappa}_i$ (see Fig.~ \ref{fig6} (c)). 

In addition, as in the mobile phone network, we find a large assortativity not only in the social connectivity, but also and more importantly in social dynamical strategies, i.e. individuals with low $\gamma$ (social keepers) tend to gather in the social network, while social explorers tend to interact between them (see Fig.~\ref{fig7SI}).
Our results show that the dynamical strategies of communication between users through Facebook wall also follow the same pattern as in mobile phone.

\begin{figure*}
\begin{center}
\includegraphics[width=\textwidth]{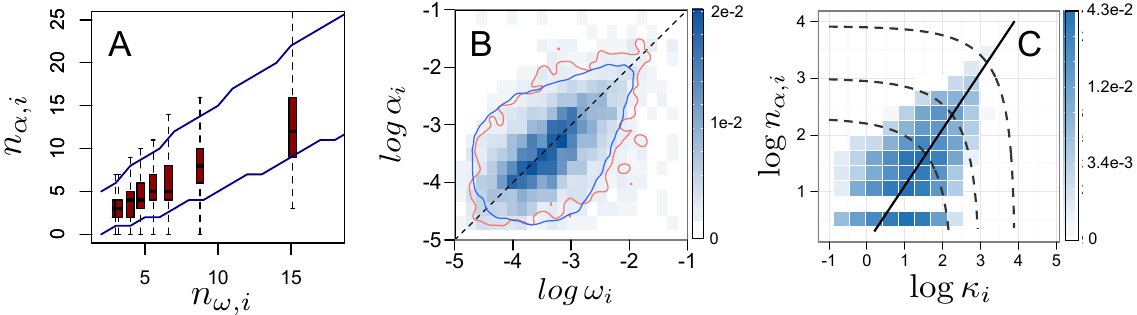}
\end{center}
\caption{Social dynamics in the Facebook database. (a) Relationship between the number of formed $n_{\alpha,i}$ and decayed $n_{\omega,i}$ ties in the observation window for the users in our database. The box plot shows the 25\% and 75\% percentiles (filled box) and 5\% and 95\% percentiles (whiskers), the solid black line is the relationship $n_{\alpha,i} = n_{\omega,i}$ and the blue curves correspond to the 5\% and 95\% percentiles of the corresponding Poisson null model in SI section E for our data. (b) Density plot $\rho(\omega_i,\alpha_i$) for the users with more than 2 ties formed and decayed. Dashed line is the $\alpha_i = \omega_i$ and the curves correspond to the contour lines $\rho = 0.03$ for the density of actual values of the rates (red) and the ones obtained in the Poissonian model in SI section E (blue). (c) Log-density plot of the social activity $n_{\alpha,i}$ and the social capacity $\overline{\kappa}_i$. Dashed lines correspond to the iso-connectivity lines $k_i(T) = 10,20,50$ and the solid line is the relationship $n_{\alpha,i} = 1.04 \overline{\kappa}_i$ obtained through PCA that explains 81\% of the variance.\label{fig6SI}}
\end{figure*}

\begin{figure*}
\begin{center}
\includegraphics[width=0.8\textwidth]{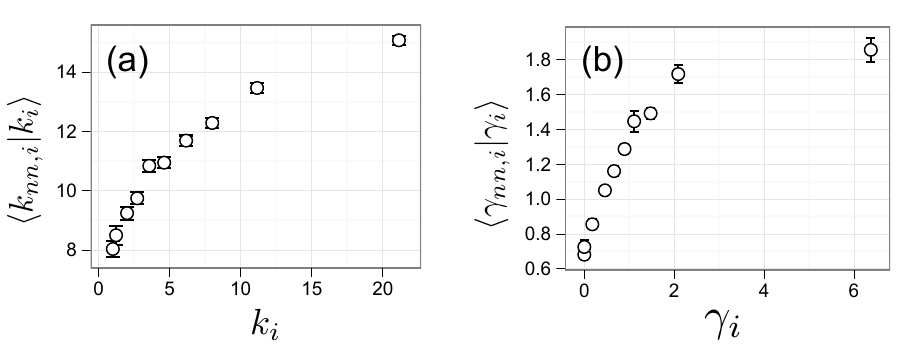}
\end{center}
\caption{Assortativity of connectivity and social strategy in Facebook social network. (a) Average next neighbor connectivity of a node $k_{nn,i}$ as a function her own connectivity $k_i$ for the $10^4$ users in the Facebook datase. A clear grown can be seen, signaling an assortativity in this social network, with a Pearson correlation coefficient $\rho(k_i,k_{nn,i})=0.257$ with confidence range [0.238,0.275]. (b) Average value of the parameter $\gamma_i$ for the neighbors of an individual as a function of her own value of $\gamma_i$. Similarly to $k_i$ we observe a clear growth and a Pearson correlation coefficient $\rho(\gamma_i,\gamma_{nn,i})=0.197$ with confidence range [0.177,0.217]. which indicates a strong assortativity of the social dynamical strategies in the Facebook database.\label{fig7SI}}
\end{figure*}

\end{document}